# Title: Skin optical attenuation coefficient modulation with arterial pulsation

Matheus Bigatão Martinelli, Christian Tolentino Dominguez, Luciano Bachmann, and George C. Cardoso[*]
Departamento de Física, FFCLRP, Universidade de São Paulo, Av. Bandeirantes 3900, Ribeirão Preto, SP 14040-901, Brazil

Keywords: Optical coherence tomography, Medical and biological imaging, Blood or tissue constituent monitoring

Abstract: Video plethysmographic (vPPG) signals arise from subtle color modulations in reflected light. vPPG signals can be used to monitor vital signs remotely in a noninvasive manner. We have used optical coherence tomography (OCT) imaging to find whether vPPG signals result from skin optical properties being modulated by arterial transmural pressure propagation. Our results confirm this hypothesis and show that the attenuation coefficient of even the most external, non-vascularized skin, changes at the same frequency as the arterial pulsation.

*Corresponding Author: E-mail: gcc@usp.br

## 1. Introduction

Video Photoplethysmography (vPPG) allows the use of visible and near-infrared (NIR) light to obtain remote plethysmographic signals in a non-invasive manner for monitoring or diagnostic purposes [1][2]. The strongest signal for visible vPPG is achieved with green light, whose penetration depth (1/e) in the skin is of the order of 0.1 mm [3]. For the inner forearm, the skin thickness from the outer stratum corneum down to the non-vascularized epidermis can be well over 70 μm [4]. However, capillary blood in the dermis starts to be significant only for depths of about 0.5 mm. Arteries are located deeper than several millimeters from the skin surface. In this scenario, Kamshilin et al. [5] estimated that periodic changes in blood volume caused by arterial pulsation (volumetric model) cannot explain vPPG signals. In view of these results, a new theory was proposed: pulse oscillations of the arterial transmural pressure modify the dermis structure, leading to periodic alterations in its optical properties.

More recently, Moço et al. [6] showed that the volumetric model is favored by results from volunteer skin experiments and Monte Carlo simulations. These authors argued that Kamshilin´s model cannot explain PPG-based oxygen saturation (SpO$_2$) measurements because skin compression due to venous pulsations would severely contaminate PPG-based SpO$_2$ readings. These conclusions are supported by (i) VIS and IR PPG measurements showing that green and IR signals come from different skin depths – the green signal comes from dermal arterioles, whereas the IR signal reaches deep subcutaneous layers; (ii) videocapillaroscopy examinations showing that the PPG signal amplitudes at green wavelengths depend on blood concentration and not on capillary density; (iii) similar PPG signal spectral dependence on the skin both under normal conditions and under compression in the range of 475 to 975 nm; and (iv) verification (by Monte Carlo simulations for multilayer normal and compressed skin models) that green light reaches pulsating arterioles at capillary loop layers at levels that are high enough to produce PPG signals both in normal and compressed skin.

The volumetric model was experimentally studied by simulating blood flow into flexible and rigid tubes of millimetric diameters [7][8]. In thick samples, the presence of a significant number of blood cells should lead to blood flow fluctuations to modify the optical properties of the samples. Drag forces should cause blood cells to rearrange, which should reorient these cells in the flow direction and consequently alter scattering and absorption cross-sections[7].

Despite much interest in the topic of vPPG [5][6][9]–[11], controversy remains, and no clear evidence of its origin has been established. vPPG signals vary significantly from region to region for no clear reason. Therefore, there is a lingering need to establish whether vPPG signals are due to changes in the epidermis optical attenuation coefficient, which in turn would be prompted by mechanical perturbations to the epidermis during the cardiac cycle. Such mechanical changes could be elicited by the arterial blood pulsation hydraulic shock caused at the artery and arteriole levels and by intrinsic arteriole and capillary compression.

OCT (Optical Coherence Tomography) is a tridimensional imaging technique that allows real-time *in vivo* visualization of a tissue's internal structures with micrometric resolution [13]. Cross-sectional tissue images (B-scans) generated by OCT enable one to monitor local structural alterations and to quantify light attenuation at different depths, to distinguish between normal and tumoral tissues, for example [12]. OCT can analyze optical properties of very superficial layers as thin as the outer skin surface stratum corneum (10–40 μm). In this way, OCT allows one to measure skin regions where no arterioles or capillary loops exist. In OCT measurement, there is no mechanical contact between the tissue and the equipment, so this imaging technique is suitable to monitor real-time modifications in the skin optical properties.

In this paper, we use OCT to study skin attenuation coefficients as a function of time to verify whether the vPPG signals could originate from periodic changes in the optical properties of the most external skin layers, outside the region where arterioles exist.

## 2. Experimental methodology

*OCT technique*

Whilst most optical diagnosis methods use incoherent light (either visible or in the NIR), OCT only detects coherently backscattered light, which is the fraction of light that undergoes constructive interference. The need for coherence limits the measurable depth in the tissue to a few hundred microns, which allows detailed monitoring of a sample's optical parameters. In fact, an OCT B-scan is built by scanning a thin slice of a thickness of the order of the OCT lateral resolution, which reduces image artifacts such as speckles and light reflections coming from dark sample regions.

For this study, OCT cross-sectional (B-scans) images of the inner left forearm skin of three volunteers were collected; the volunteers had the forearm held in place by styrofoam molds. The B-scans were acquired by a swept source optical coherence tomography (SS-OCT) imaging system (model OCS1300SS, Thorlabs Inc.) at a rate of 10 images/s for 30 s. The spatial dimensions of each image were 5 mm x 3 mm (width x depth), and the pixel resolution was 1024 x 512 for width and depth, respectively. The main characteristics of the OCT equipment were sweep laser source centered at 1300 nm, average output power of 10 mW, axial scan rate of 16 kHz, and axial and lateral imaging resolutions of 12 and 25 µm, respectively. The laser beam was focused on the sample with a 25-mm objective lens, and a pair of XY galvo mirrors scanned the laser beam across the sample surface. The coherent backscattering signal was detected and processed by the OCT system.

A finger photoplethysmographer (model CMS 50D+ Blue Finger Pulse Oximeter with USB) simultaneously recorded the photoplethysmogram from the index finger of the same arm under OCT measurements. The room was kept at 23 °C throughout the experiment.

The investigation conformed to the principles in the Declaration of Helsinki (World Medical Association 2000) and all subjects gave their written informed consent.

*Signal Processing*

Light traveling through homogeneous and low-scattering samples follows an exponential decay according to $I(z) \propto [\exp(-2\mu z)]^{1/2}$, where $\mu$ is the attenuation coefficient, $I(z)$ is the optical signal amplitude, and $z$ is the penetration depth [13]. The factor 2 multiplying $\mu$ accounts for the round trip and is compensated by the square root, leading to the ordinary Beer-Lambert law $I(z) \propto \exp(-\mu z)$. The square root comes from the fact that the OCT signal originates from light interference, which is proportional to the coherent light E-field. The attenuation coefficient $\mu$ takes both the absorption coefficient and the reduced scattering coefficient into account. The reduced scattering coefficient also depends on the sample anisotropy factor (g).

For image analysis, the central region of the OCT B-scan was selected as a region of interest (ROI). This ROI had width of $\Delta x = 600$ μm and depth of $\Delta z = 2.00$ mm, with $x_o = 0$ at the image horizontal symmetry center and $z_o = 0$ at the air/skin interface (see Fig.1). Keeping the ROI near $x_o$ is important to minimize the OCT laser beam angular deviations and to reduce the uncertainty regarding $\mu$ determination.

The OCT images were analyzed by a script written in MATLAB v. 2016b (Mathworks, USA). For each B-scan image, the skin outer surface at position $x = 0$ was automatically defined by the script as $z = 0$. Next, for each B-scan, the average A-scan profile along the x-direction of the ROI was produced and normalized by its maximum intensity. The inset in Figure 1 illustrates a typical averaged A-scan profile. At an acquisition rate of 10 Hz, ten of such curves are produced per second. A Beer-Lambert curve is adjusted to each averaged A-scan. Because the OCT B-scan image is given in decibels (dB), the Beer-lambert law becomes linear (Figure 1, inset). For depths above 0.5 mm, the image becomes too contaminated with OCT background noise.

## 3. Results

Figure 1 shows a B-scan image of the skin inner forearm region investigated herein. The decay curve in Figure 1, inset, is the OCT optical interference signal amplitude as a function of skin penetration, as described above.

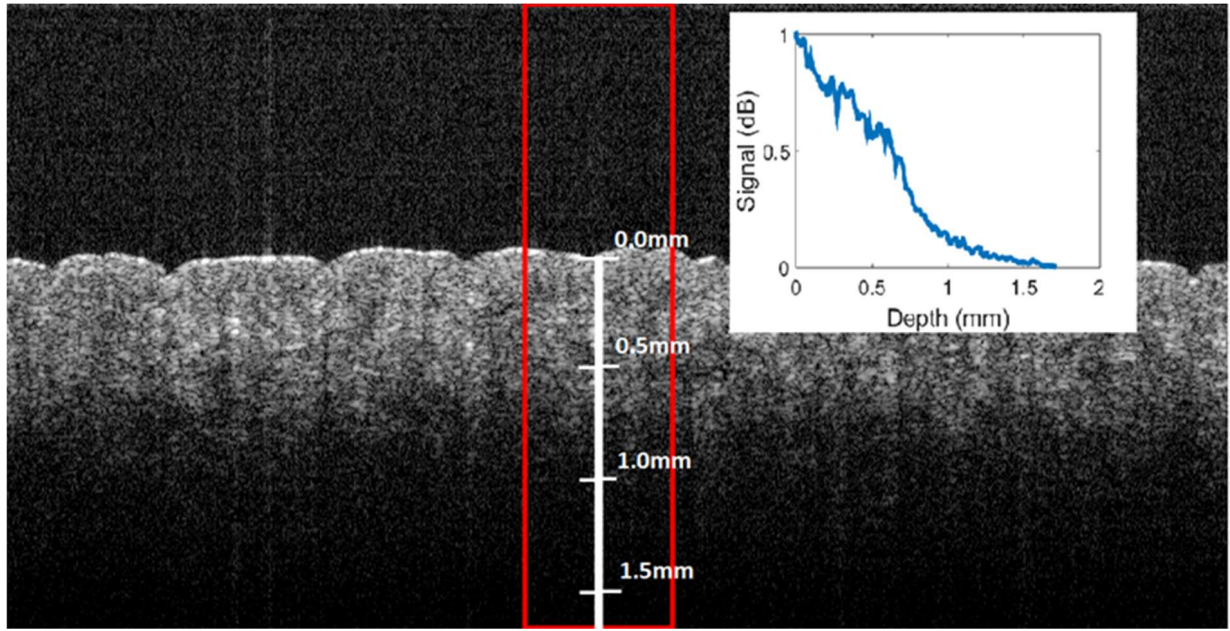

Figure 1. Inner forearm skin OCT B-scan image. The inset is the OCT optical interference signal in dB as a function of depth starting at the skin surface, averaged over the width ($\Delta x = 600$ μm) of the ROI delimited by the red rectangle.

We performed linear fits for the region spanning from $z = 0.0$ to $z = 0.5$ mm in curves such as the one depicted in Figure 1, upper inset. The attenuation coefficient μ is the slope of such a fit, which we determined for each B-scan. We used this procedure to obtain the attenuation coefficient time dependence and acquired the B-scans at a rate of 10 Hz. Figure 2(a) displays the fast Fourier transform (FFT) of μ(t) calculated from OCT ROI data, whereas Figure 2(b) shows the FFT of the simultaneously acquired finger PPG signal (A/D conversion rate of 60 Hz). In this instance, observe that the heart rate frequency (HR) of 72 beats per minute (bpm) clearly appears in both curves, indicating that the skin attenuation coefficient changes with HR. We found equivalent results for the other two individuals for HR varying between 58 and 72 bpm. To observe the HR in the FFT with good SNR, it is essential that the heart rate variation (HRV) remain small during the 30-second measurement. When HR varied during acquisition, we discarded the measurements because their FFT peaks were not well defined.

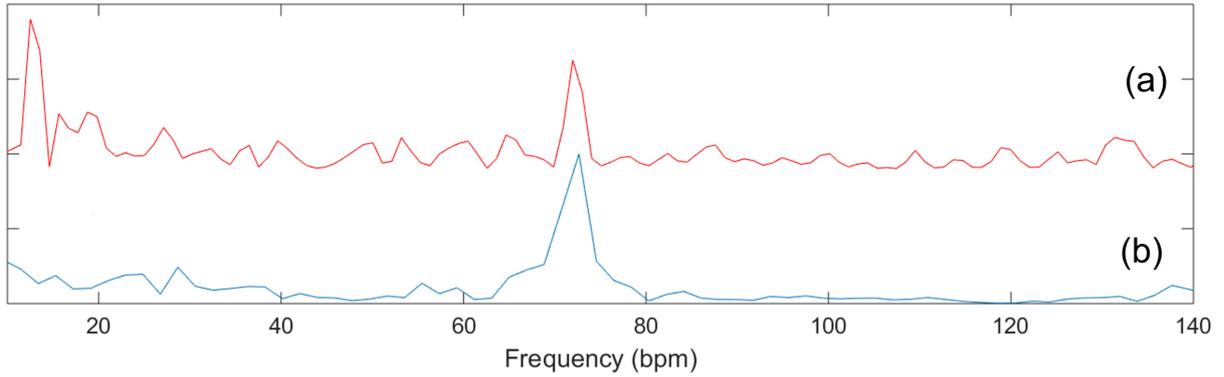

Figure 2. Power spectral density for (a) skin attenuation coefficient µ and (b) finger PPG signal, for simultaneous signal acquisition. The peak at 72 bpm is the heart rate frequency.

Finally, to verify whether arm ballistocardiac vibrations during the experiment could explain the peak at HR in FFT(µ(t)), we compared spatial with temporal variation in µ(x,t). If the spatial µ inhomogeneity is significant as compared to the µ variation in time, ballistocardiac vibrations could explain the µ fluctuation observed for HR. To determine how the attenuation coefficient varies with the x-position (the central 40% of the OCT images, or 2 mm), we time-averaged µ along the 30-second time window, to obtain $\langle\mu(x)\rangle_T$, as shown in Figure 3(a). Subindex T represents averaging in time. To determine $\langle\mu(x)\rangle_T$, we divided each B-scan into 20-µm wide A-scan slices and calculated the attenuation coefficient for each slice. To determine the corresponding temporal dependency $\langle\mu(t)\rangle_X$, we performed an analogous calculation over the different 20-µm slices, as shown in Figure 3b. Notice that $\langle\mu(t)\rangle_X$ here uses a wider X-width than the signal used for FFTs of the type shown in Figure 2(a), where we used an X-width of 600 µm instead. The standard deviation of spatial and temporal dependences ( $\langle\mu(x)\rangle_T$ and $\langle\mu(t)\rangle_X$ ) are $\sigma_{\mu(x)} = 0.096$ mm$^{-1}$ and $\sigma_{\mu(t)} = 0.23$ mm$^{-1}$, respectively, whilst the overall average value for the attenuation coefficient $\langle\mu\rangle$ is $(1.2 \pm 0.1)$ mm$^{-1}$. Although we did not aim to measure the attenuation coefficient, our result agrees with the values reported by Su et al. [14], who obtained µ ranging from 0.7 to 1.2 mm$^{-1}$ for *in vivo* forearm skin layers at depths ranging from 200 to 450 µm.

The results above show that the spatial change in the attenuation coefficient is smaller than its variation in the temporal dimension. Therefore, there is an actual temporal variation in the attenuation coefficient at HR frequency; that is, the variation is not simply an artifact due to spatial skin inhomogeneities and minor skin motion at heart rate.

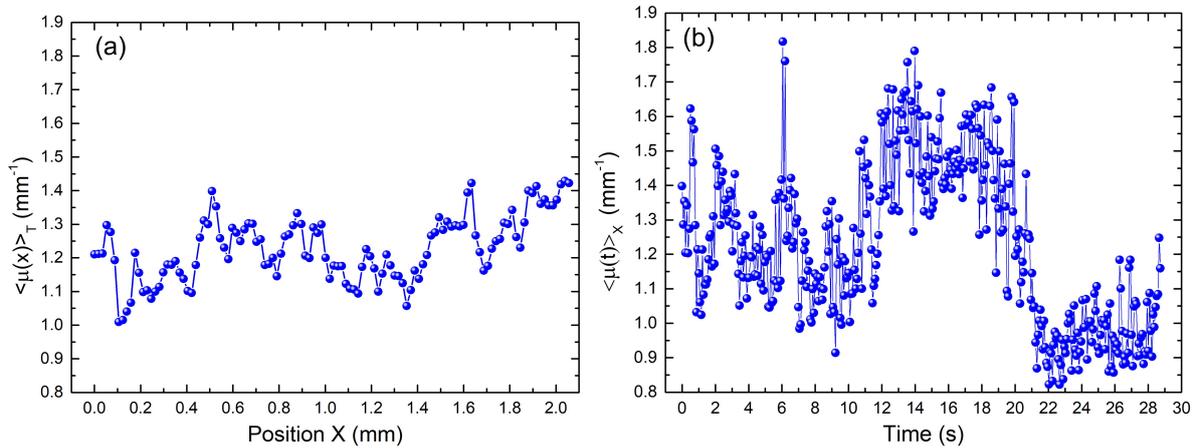

Figure 3. (a) Spatial and (b) temporal skin attenuation coefficient variations.

4. Discussion

The results shown in Figure 2 originate from the modified skin optical properties arising from mechanical pulse propagation from the arteries toward the tissue lateral layers. Kamshilin et al. [5][15][16] suggested this mechanism to explain their vPPG results. By using green light (525 nm), these authors observed PPG signals coming from a subject's skin, with or without mechanical contact. Capillary time dynamics is much slower than the heart rate and cannot explain the results observed. The authors attributed their observations to skin deformations provoked by mechanical forces generated during arterial pulsations, which would modify the outer skin layer absorption and scattering coefficients. For example, Kirillin et al. [17] analyzed differences between OCT images acquired under low (~0.07 N/mm$^2$) and higher (> 0.35 N/mm$^2$) pressure and verified that the skin optical attenuation properties changed significantly. The latter study concluded that OCT imaging could measure structural modifications in skin layers as a consequence of applied compression forces.

Red and near-IR light penetrates tissues for several millimeters [18]–[20]. In our experiments, we observed that OCT light at 1300 nm penetrates around 0.8 mm (1/μ), which meets our purpose of surveying the most external skin structures, as shown in Figure 1. Up to around 300 μm, not even capillaries exist. Further down, blood capillaries and lymphatic flux might occur. Veins and arterioles are located several millimeters away from the skin surface [1][21] and do not contribute to OCT signals; they are only responsible for communicating blood pressure toward outer skin layers. We hypothesized that changes in the attenuation coefficient at the cardiac frequency are due to a blood pressure hydraulic shock effect on the capillaries and ensuing tissues [22]. Because the capillary flow is slow as compared to the cardiac cycle dynamics [23]–[25], this flow does not influence our results. Therefore, we inferred that alterations in the skin optical properties only stem from modifications in the skin internal structure, which result from propagation of mechanical compression pulses coming from internal blood vessels toward the outer skin layers. Hence, the vPPG signal depends both on the position of the underlying arteries and on the mechanical properties of the surrounding skin layers, explaining why the vPPG intensity varies according to the skin region [28].

We ruled out ballistocardiac effects by showing that temporal attenuation coefficient variations are significantly more relevant than spatial variations. This agrees with recent findings that vPPG does not have a ballistocardiac origin [26][27]

Our results corroborate with the vPPG mechanism suggested by Kamshilin et al. [5] (who indicated that skin compression causes the vPPG signal) without disagreeing with the experimental observations of Moço et al. [6] mentioned in our introductory section. However, our results do not corroborate with the volumetric model; that is, on the basis of our results, periodic changes in perfusion or blood flow in the outer skin investigated here are not necessary for vPPG signal acquisition.

The fact that vPPG might be related to a hydraulic shock effect [22] due to blood pulsation reflection from arteries into capillaries does not preclude that physiological changes or contractions of capillaries and arterioles also contribute to alterations in optical properties. Such effects could cause periodic modifications in the tissue, which would not necessarily be in phase with blood pulsation and could enhance or decrease the overall vPPG signal. Also, our findings do not explain why green light presents the strongest vPPG signal. We believe that the heightened blood absorption at these wavelengths will make any light modulation more prominent, since tissues contain a significant amount of blood.

The fact that we conducted the experiments with light at 1300 nm does not affect our results because changes in optical attenuation coefficient are surrogates for skin mechanical alterations.

Our results show that changes in superficial skin optical properties occur at cardiac frequency without the need for direct contribution from blood flow. This finding explains why remote reflective vPPG can be conducted by employing wavelengths that do not penetrate significantly, such as blue light [1][28]–[31]. The suggestion that the vPPG signal originates from propagation of arterial transmural pulsation implies that vPPG is also a function of the surrounding tissue local viscoelastic properties, with possible clinical applications that would require further studies.

4. Conclusion

We have determined that the attenuation coefficient power spectrum of the epidermis outer layer contains the cardiac frequency. The observed optical attenuation coefficient changes are surrogate for modulation of its mechanical properties. Our finding helps to understand why visible light and vPPG can be used for remotely obtaining HR in skin regions with very low vascularization. In particular, our finding explains why vPPG can be conducted with blue light, which does not penetrate further than a few hundred microns. We speculate that changes in the attenuation coefficient at such shallow depths are due to a hydraulic shock effect, where blood pulsation in deeper skin layers is propagated through capillaries and eventually through tissues, causing a mechanical shock that modifies the mechanical and therefore the optical properties of the skin outer layer. Consequently, the location of the underlying arteries and the tissue local viscoelastic properties can affect the vPPG signal amplitude.

**Acknowledgments**

The authors gratefully acknowledge Prof. Jean Pierre von de Weid and Centro de Estudos em Telecomunicações (CETUC) of Pontifícia Universidade Católica do Rio de Janeiro for providing OCT equipment and software. This work was supported by Coordenação de Aperfeiçoamento de Pessoal de Nível Superior-CAPES, São Paulo Research Foundation (FAPESP) grants #2017/01932-5 and 2013/07276-1 (FAPESP, CEPOF CEPID).